\def\ls{\lower4pt\hbox{${\buildrel < \over \sim}$}}
\def\gs{\lower4pt\hbox{${\buildrel > \over \sim}$}}

\documentclass[12pt, preprint]{aastex}

\slugcomment{Accepted for publication in {\it The Astrophysical Journal}}

\shorttitle{Optical Monitoring of 1ES 1011+496}
\shortauthors{B\"ottcher et al.}

\begin{document}

\title{Optical Spectral Variability of the Very-High-Energy Gamma-Ray Blazar 
1ES 1011+496}

\author{M. B\"ottcher\altaffilmark{1}, 
B. Hivick\altaffilmark{1}, 
J. Dashti\altaffilmark{1}, 
K. Fultz\altaffilmark{1},
S. Gupta\altaffilmark{1}, 
C. Gusbar\altaffilmark{1},
M. Joshi\altaffilmark{1,2},
A. Lamerato\altaffilmark{1}, 
T. Peery\altaffilmark{1},
D. Principe\altaffilmark{1,3},
A. Rajasingam\altaffilmark{1},
P. Roustazadeh\altaffilmark{1},
J. Shields\altaffilmark{1}
}

\altaffiltext{1}{Astrophysical Institute, Department of Physics and Astronomy, \\
Clippinger 251B, Ohio University, Athens, OH 45701, USA}
\altaffiltext{2}{Institute for Astrophysical Research, Boston University, \\
725 Commonwealth Ave., Boston, MA 02215, USA}
\altaffiltext{3}{Rochester Institute of Technology, 84 Lomb Memorial Drive, \\
Rochester, NY 14623, USA}

\begin{abstract}
We present results of five years of optical (UBVRI) observations of the 
very-high-energy gamma-ray blazar 1ES~1011+496 at the MDM Observatory. 
We calibrated UBVRI magnitudes of five comparison stars in the field of 
the object. Most of our observations were done during moderately faint 
states of 1ES~1011+496 with $R \gtrsim 15.0$. The light curves exhibit 
moderate, closely correlated variability in all optical wavebands on 
time scales of a few days. A cross-correlation analysis 
between optical bands does not show significant evidence for time lags. 
We find a positive correlation (Pearson's $r = 0.57$; probability
of non-correlation $P(>r) \approx 4 \times 10^{-8}$)
between the R-band 
magnitude and the B - R color index, indicating a bluer-when-brighter
trend. Snap-shot optical spectral energy distributions (SEDs) exhibit a peak 
within the optical regime, typically between the V and B bands. We find a 
strong ($r = 0.78$; 
probability of non-correlation $P (>r) \approx 10^{-15}$)
positive correlation between the $\nu F_{\nu}$ peak flux 
and the peak frequency, best fit by a relation $\nu F_{\nu}^{\rm pk} \propto 
\nu_{\rm pk}^k$ with $k = 2.05 \pm 0.17$. Such a correlation is consistent
with the optical (synchrotron) variability of 1ES~1011+496 being primarily 
driven by changes in the magnetic field.
\end{abstract}


\section{Introduction}

Blazars are the most violent class of active galactic nuclei, consisting
of flat-spectrum radio quasars (FSRQs) and BL~Lac objects 
(named after their historical prototype, BL~Lacertae). 
They exhibit
rapid variability down to time scales as short as a few minutes 
\citep{aharonian07,albert07a}. Their observed flux is dominated by a
non-thermal continuum exhibiting two broad spectral bumps: A low-frequency
bump from radio to UV -- X-ray frequencies, and a high-frequency component
from X-ray to $\gamma$-rays. 
In the framework of relativistic jet models, the low-frequency (radio
-- optical/UV) emission from blazars is interpreted as synchrotron
emission from nonthermal electrons in a relativistic jet. The
high-frequency (X-ray -- $\gamma$-ray) emission could either be
produced via Compton upscattering of low frequency radiation by the
same electrons responsible for the synchrotron emission \citep[leptonic
jet models; for a recent review see, e.g.,][]{boettcher07}, or 
due to hadronic processes initiated by relativistic protons 
co-accelerated with the electrons \citep[hadronic models, for 
a recent discussion see, e.g.,][]{muecke01,muecke03}. 

To date, about 30 blazars have been detected in very high energy (VHE, 
$> 100$~GeV) $\gamma$-rays with ground-based air \v Cerenkov telescope 
facilities\footnote{For a complete list of VHE $\gamma$-ray sources 
see {\tt http://tevcat.uchicago.edu}}. Most of these TeV blazars belong 
to the sub-class of high-frequency peaked BL~Lac objects (HBLs). They 
are characterized by a synchrotron spectrum peaking at frequencies 
$\nu_{\rm sy}^{\rm pk} \gtrsim 10^{15}$~Hz, i.e., in the UV or X-ray 
range, and $\gamma$-ray peaks at $\nu_{\gamma}^{\rm pk} \gtrsim 
10^{25}$~Hz, i.e., typically beyond the Fermi energy range \citep{abdo10}. 
VHE $\gamma$-rays from sources at cosmological distances can be absorbed 
by the extragalactic background light (EBL) due to $\gamma\gamma$ pair 
production \citep[e.g.,][]{dk05,stecker06,franceschini08,gilmore09,finke10}.
Hence, the EBL absorption of distant VHE $\gamma$-ray sources may provide
a probe of the spectrum and cosmological evolution of the EBL, which is
notoriously difficult to measure directly due to bright foregrounds. 

The BL~Lac object 1ES~1011+496 was detected as a VHE $\gamma$-ray emitter
by MAGIC in the spring of 2007 \citep{albert07b}. Follow-up optical spectroscopy
at the MMT confirmed the previously uncertain redshift of $z = 0.212 \pm 0.002$
for this source \citep{albert07b}. At the time of its VHE detection, 1ES~1011+496 
was the most distant VHE $\gamma$-ray source known with a well-determined
redshift, and to date it still ranks among the top five. It might therefore
offer a prime opportunity for studying the EBL through its absorption signature 
at VHE $\gamma$-rays. However, in order to exploit this opportunity, a thorough 
understanding of the intrinsic spectral energy distribution (SED) of the source, 
constrained through observations at lower (radio through GeV $\gamma$-ray) 
frequencies, is essential. For this purpose, we present here a detailed 
study of the optical spectral variability of this object with data gathered
over the course of 5 years at the 1.3m McGraw-Hill Telescope of the MDM
Observatory on Kitt Peak, Arizona.

The VHE $\gamma$-ray detection of 1ES~1011+496 was triggered by a large
optical outburst of the source in 2007 March \citep{albert07b}. This object
is regularly monitored in the optical R-band through the Turku blazar
monitoring program led by Kari Nilsson, with the 1.03~m telescope of 
the Tuorla Observatory in Finland, as well as the 35~cm telescope of the
KVA Observatory on La Palma, Canary Islands, Spain\footnote{Daily light
curves are posted at {\tt http://users.utu.fi/kani/1m/index.html}}. 
Apart from this, 1ES~1011+496 has so far received rather little attention
by optical observers, and its optical spectral variability has remained
unexplored. The object has been observed in X-rays by Einstein \citep{elvis92} 
and more recently, Swift/XRT \citep{abdo10}. In all observations, it shows a 
steep X-ray spectrum, indicating the dominance of synchrotron emission
in the X-ray regime. 1ES~1011+496 may also be associated with the EGRET 
$\gamma$-ray source 3EG~J1009+4855 \citep{hartman99}, although this
association is uncertain \citep{sowards03}. The object is clearly 
detected by Fermi and listed in the Fermi 3-month catalogue as the 
source 0FGL~J1015.2+4927. The Fermi data reveal a rising $\nu F_{\nu}$ 
spectrum (i.e., photon index $\Gamma < 2$) in the 100~MeV -- 30~GeV 
energy range \citep{abdo10}. 

Optical/UV observations were performed by Swift/UVOT in May 2008
\citep{abdo10}, 
during
the rising phase of an optical outburst similar to the one triggering the 
MAGIC discovery observations in 2007. During those observations, Swift/UVOT
measured a rising $\nu F_{\nu}$ optical/UV continuum spectrum. This, together 
with the hard Fermi spectrum, justified the classification of 1ES~1011+496 
as an HBL.

\begin{figure}
\plotone{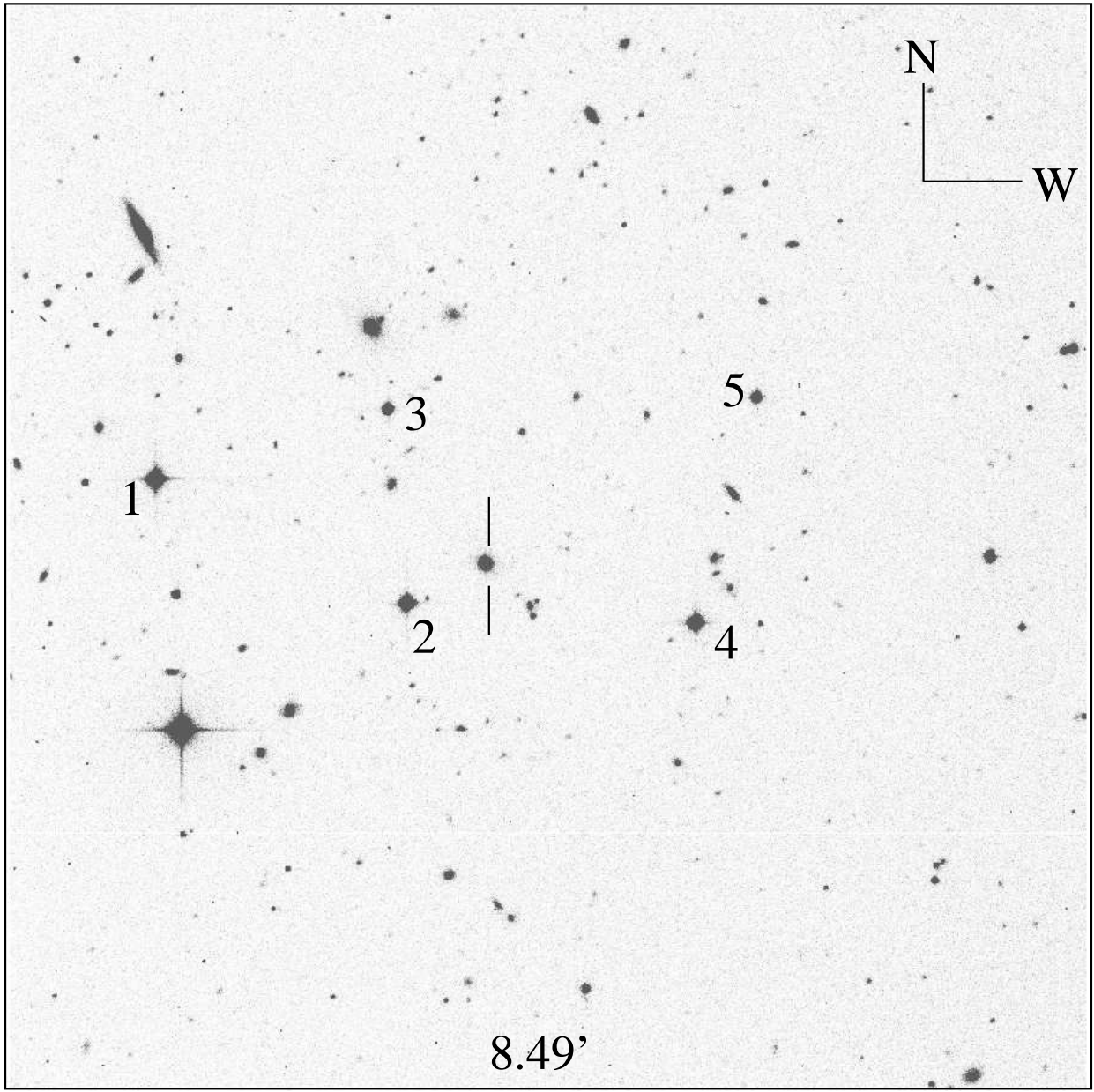}
\caption{Finding chart (R band) of the field around 1ES~1011+496 with 5 
comparison stars}
\label{findingchart}
\end{figure}

Over the course of $\sim 5$~years, we have collected optical multi-band
(UBVRI) photometric data on 1ES~1011+496. As there are no comparison stars 
with reliably calibrated UBVRI magnitudes in the field of view of this 
object, we calibrated the magnitudes for 5 comparison stars. In \S 
\ref{observations} we describe the observations, data reduction, the 
calibration of comparison-star magnitudes, as well as general features
of the light curves. Results of a spectral-variability study are
presented in \S \ref{variability}. We performed a cross-correlation
analysis between the variability patterns in different optical bands,
which we describe in \S \ref{crosscorrelations}. We summarize and
discuss our results in \S \ref{discussion}.

\section{\label{observations}Observations, data reduction, 
and light curves}

Optical (UBVRI) data were collected at the 1.3m McGraw-Hill Telescope of
the MDM Observatory on the south-west ridge of Kitt Peak, Arizona, during
9 $\sim 1$-week observing runs in 2005 April, September, and December,
2008 April and May, 2009 February and May, and 2010 April and July. The
telescope is equipped with standard Johnson-Cousins UBVRI filters. 
Exposure times for science frames on 1ES~1011+496 were between 60 -- 180~s,
depending on filter and atmospheric conditions. All frames were bias-subtracted
and flat-field corrected using standard routines in IRAF. 

\begin{deluxetable}{cccccc}
\tabletypesize{\scriptsize}
\tablecaption{Calibrated Magnitudes of Comparison Stars in Fig. 
\ref{findingchart}}
\tablewidth{0pt}
\tablehead{
\colhead{Star} & \colhead{U} & \colhead{B} & \colhead{V} & \colhead{R} &
\colhead{I}
}
\startdata
1 & $15.355 \pm 0.010$ & $14.738 \pm 0.002$ & $13.885 \pm 0.002$ & $13.493 \pm 0.002$ & $12.973 \pm 0.003$ \\
2 & $15.230 \pm 0.050$ & $15.109 \pm 0.003$ & $14.447 \pm 0.002$ & $14.069 \pm 0.002$ & $13.703 \pm 0.003$ \\
3 & $17.456 \pm 0.043$ & $16.788 \pm 0.007$ & $15.918 \pm 0.003$ & $15.437 \pm 0.005$ & $15.013 \pm 0.007$ \\
4 & $14.488 \pm 0.018$ & $14.698 \pm 0.002$ & $14.334 \pm 0.002$ & $14.067 \pm 0.002$ & $13.797 \pm 0.003$ \\
5 & $16.477 \pm 0.018$ & $16.539 \pm 0.008$ & $15.794 \pm 0.003$ & $15.448 \pm 0.005$ & $15.125 \pm 0.008$ \\
\noalign{\smallskip\hrule\smallskip}
\enddata
\label{standards}
\end{deluxetable}

\begin{figure}
\plotone{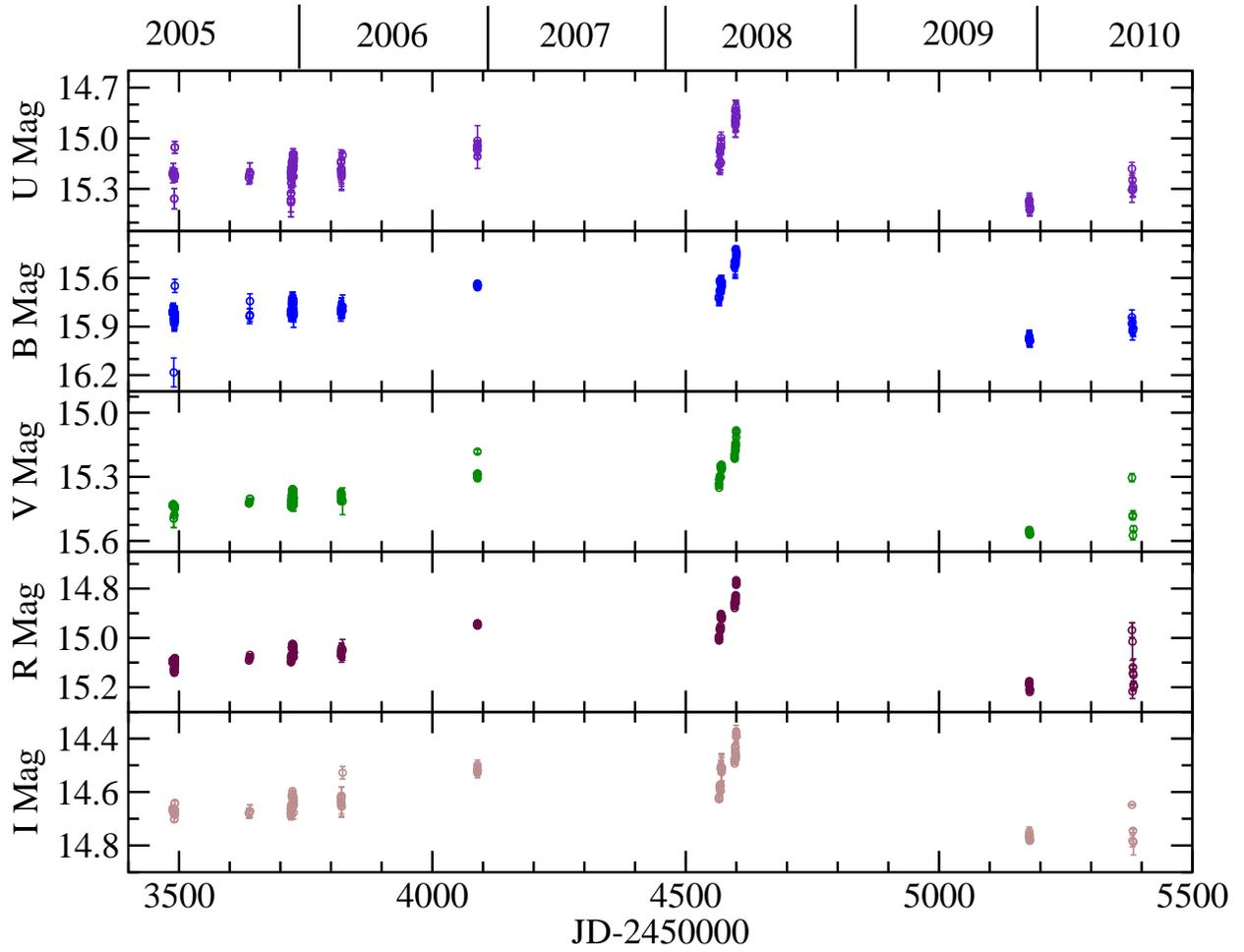}
\caption{Multi-band (UBVRI) light curves of 1ES~1011+496 from our MDM observations}
\label{lightcurves}
\end{figure}

In order to be able to perform relative photometry with comparison stars 
in the field of view, well-calibrated UBVRI magnitudes of those stars need
to be known. Since no such information is currently available in the literature,
we calibrated UBVRI magnitudes of five comparison stars in the field of view, 
as indicated in Fig. \ref{findingchart}. For the calibration, we obtained a
total of 30 (2 per filter per standard) exposures of the Landoldt Equatorial 
standards \citep{landoldt92} PG~1525-071, PG~1633+099, PG~1657+078 on the
night of April 28, 2005, which provided photometric observing conditions.
Instrumental magnitudes of these three Equatorial standards as well
as our comparison stars in the field of 1ES~1011+496 were extracted using
the {\it phot} routine within the IRAF package DAOPHOT. Following standard 
procedures for IRAF photometric calibrations \citep{massey92}, we used the
routine {\it fitparams} to solve the transformation equations to evaluate
the calibrated, physical magnitudes of our standard stars.
The resulting calibrated magnitudes are listed in Table \ref{standards}.

\begin{figure}
\plotone{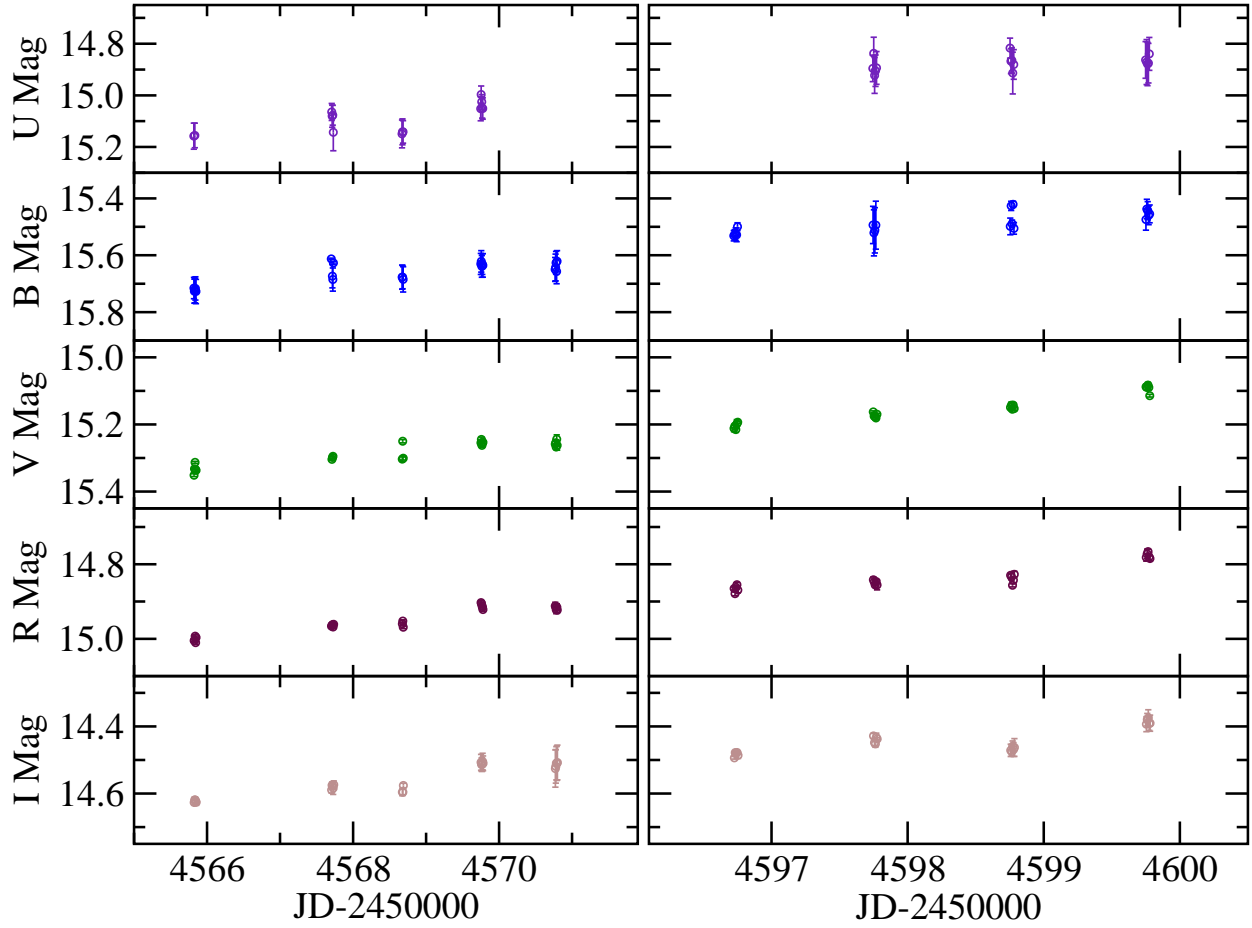}
\caption{Multi-band (UBVRI) light curves of 1ES~1011+496 during the high activity 
state in April -- May 2008}
\label{lc2008}
\end{figure}

After calibration of our comparison stars, we extracted instrumental
magnitudes of the comparison stars and 1ES~1011+496 using the {\it phot}
routine within the DAOPHOT package of IRAF, and converted instrumental
to physical magnitudes assuming that the difference between instrumental
and physical magnitudes is the same for the object and all comparison stars.
The resulting light curves for all observing runs combined are displayed
in Fig. \ref{lightcurves}. The figure illustrates that for most of our runs,
the object was in a rather faint optical state with $R \gtrsim 15.0$, and
shows very moderate variability. 

However, we did observe substantial variability during our two observing
runs in 2008 April 9 -- 14 and May 10 -- 13. This was during the rising 
phase of a major outburst that peaked later that year. Unfortunately, to
our knowledge, there are no observations available covering the peak of 
that outburst. During our observations we found a maximum R-band brightness 
of $R \sim 14.8$. The object later exceeded $R_{\rm peak} < 
14.6$\footnote{\tt http://users.utu.fi/kani/1m/index.html}. Fig. \ref{lc2008}
shows the multi-band light curves from our 2008 April and May runs. They
exhibit variability on time scales of a few days, but no evidence for
intraday variability. The variability in all optical bands appears well
correlated. This correlation will be investigated in more detail in
\S \ref{crosscorrelations}.

\begin{figure}
\plotone{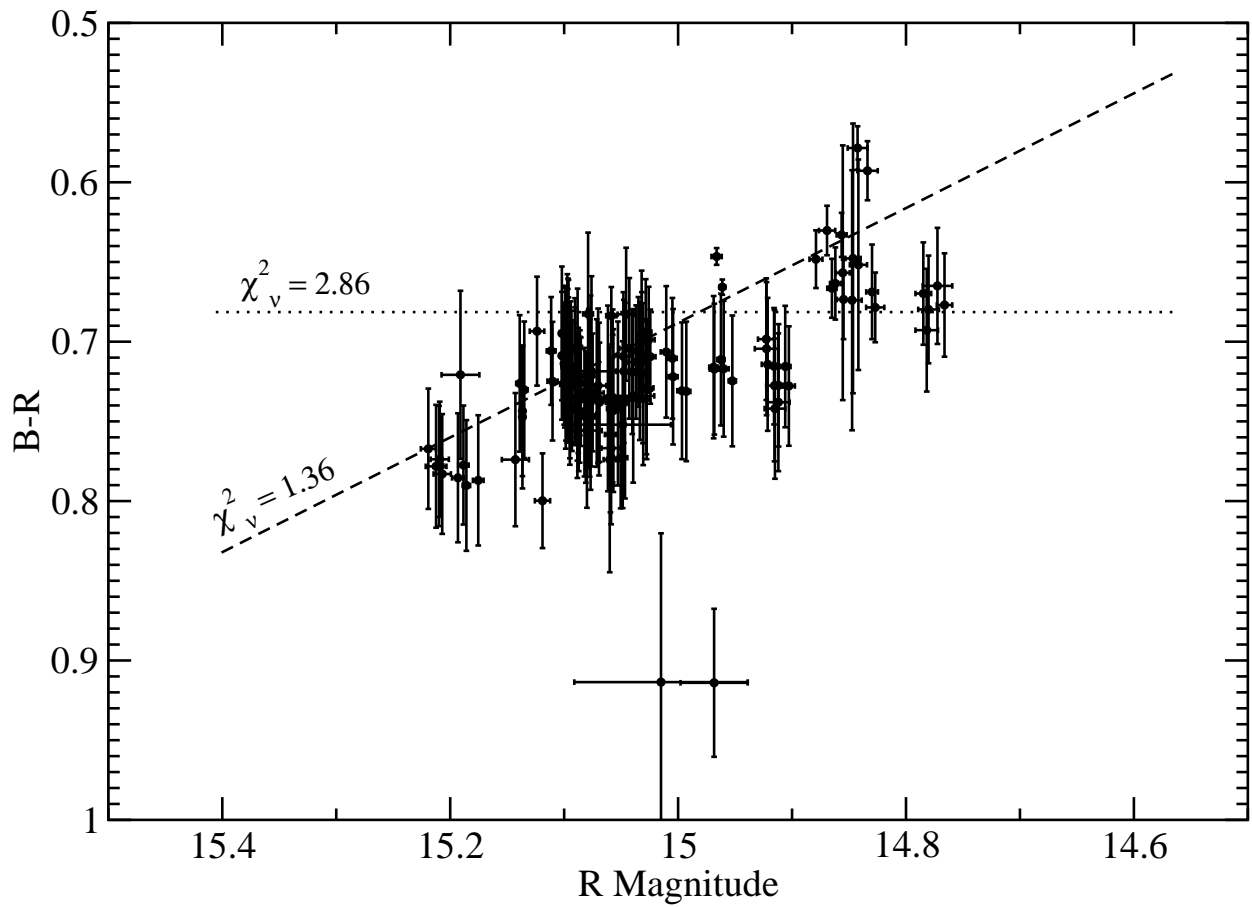}
\caption{Color-magnitude diagram for 1ES~1011+496. The data show significant
color variability. A positive linear correlation (bluer when brighter)
is indicated by a Pearson's correlation coefficient of $r = 0.57$, with
a probability for non-correlation of $P (> r) \approx 4 \times 10^{-9}$. }
\label{color}
\end{figure}

\section{\label{variability}Optical spectral variability}

In order to test whether the variability discussed in the previous section
is associated with spectral changes, we first calculated B - R color indices
for any pair of B and R magnitudes measured within 15 minutes of each other.
The resulting color-magnitude diagram (R magnitude vs. B - R color) is shown 
in Figure \ref{color}. 
Error bars on B - R are calculated via standard error progatation, i.e.,
$\sigma_{B - R} = \sqrt{\sigma_R^2 + \sigma_B^2}$.
The data clearly indicate color variability. A fit
of a constant B - R color as a function of R magnitude results in $\chi_{\nu}^2 
= 2.86$. A fit of a linear correlation results in a marginally acceptable
$\chi_{\nu}^2 = 1.36$. A correlation analysis of the color-magnitude data set
yields a Pearson's correlation coefficient of $r = 0.57$, 
which is generally interpreted as a weak positive correlation 
between color and magnitude. In order to quantify the probability 
of such a correlation coefficient resulting from an uncorrelated 
data set, we performed Monte-Carlo simulations of 1 billion randomly 
produced, uncorrelated data sets, extending over similar spreads of 
values, and with the same number of data points as our observational
data set. The Pearson's correlation coefficient for each set was evaluated, and
from the entire ensemble, the probability of a correlation coefficient 
$\vert r \vert > x$ for values of $0 < x < 1$ resulting from an uncorrelated 
data set was evaluated. From these simulations, we find that the probability 
of a correlation coefficient of $r = 0.57$ in a data set with the characteristics 
of our R vs. B - R data is $P (\vert r \vert \ge 0.57) \approx 4 \times 10^{-9}$,
indicating, in fact, a highly significant correlation. The observed correlation 
corresponds to
a bluer-when-brighter trend, as observed in most BL~Lac objects. This is likely
to reflect the dynamics of the non-thermal synchrotron emission from the jet
dominating in the optical regime. 

\begin{figure}
\plotone{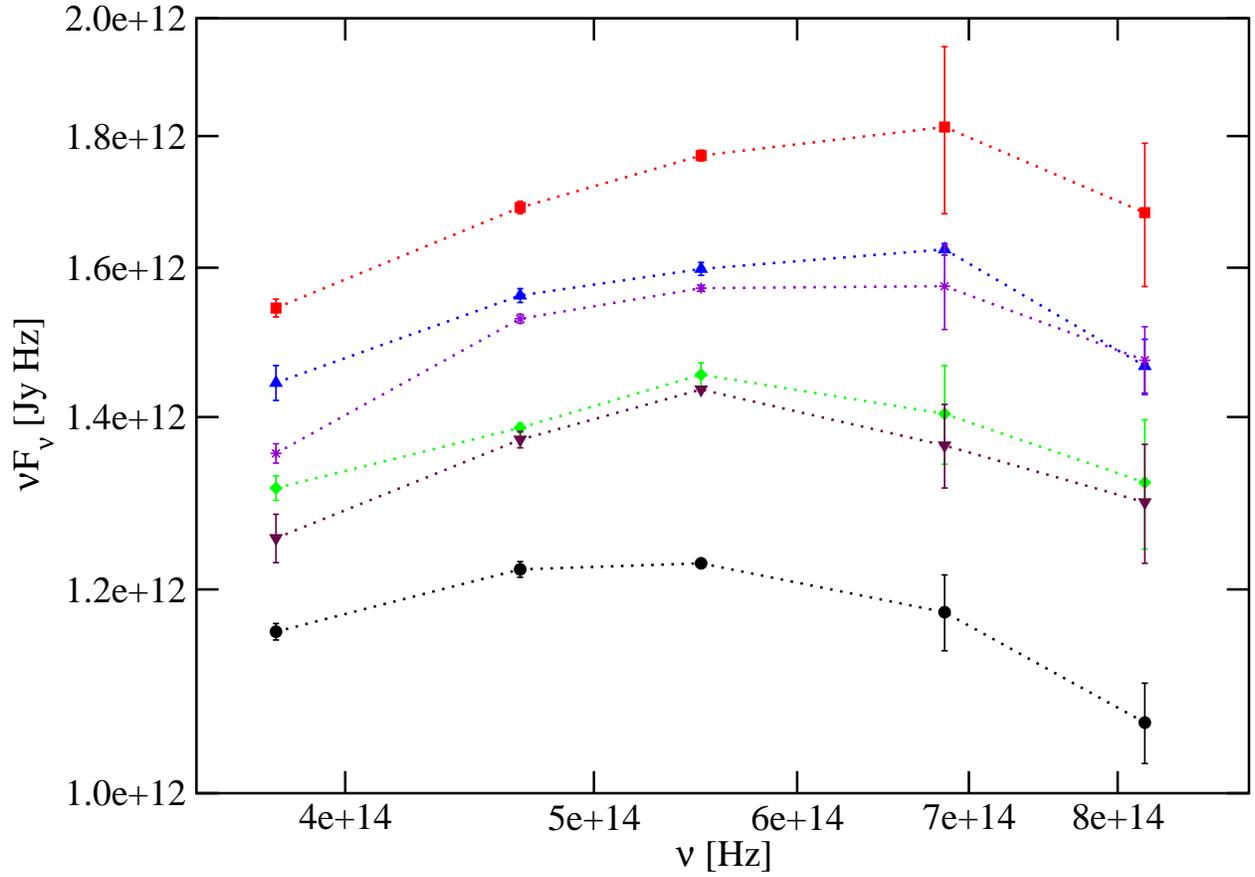}
\caption{Snap-shot UBVRI SEDs of 1ES~1011+496. The SEDs reveal a $\nu F_{\nu}$
peak typically between the B and V band and suggest a positive correlation 
between $\nu F_{\nu}$ peak flux and peak frequency.}
\label{SEDs}
\end{figure}

\begin{figure}
\plotone{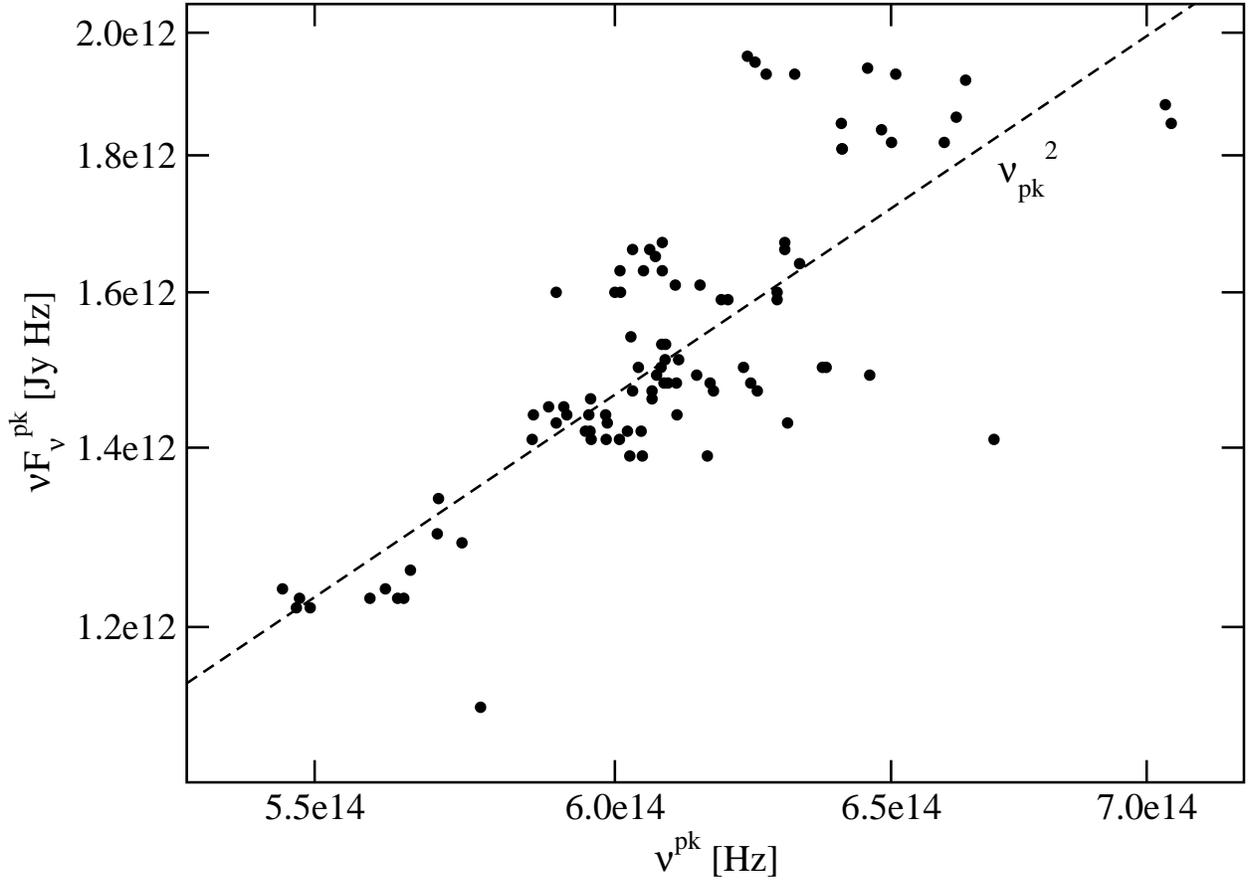}
\caption{Best-fit peak frequency vs. peak $\nu F_{\nu}$ flux for the entire
data set. The data show a significant correlation ($r = 0.78$, $P (> r) 
\approx 10^{-15}$), best fit 
by a dependence $\nu F_{\nu}^{\rm pk} \propto \nu_{\rm pk}^k$ with $k =
2.05 \pm 0.17$. The dashed line indicates a putative correlation with
index $k = 2$. }
\label{nupeak}
\end{figure}

In order to investigate spectral changes in the optical continuum in more
detail, we extracted UBVRI SEDs for all sequences of magnitudes measured
within 15 minutes of each other. For this purpose, the magnitudes were 
de-reddened using the Galactic extinction coefficients as given in the NASA 
Extragalactic Database\footnote{\tt http://nedwww.ipac.caltech.edu/}
and converted to $\nu F_{\nu}$ fluxes. A representative sample of the 
resulting SEDs is plotted in Figure \ref{SEDs}. The SEDs all exhibit
a $\nu F_{\nu}$ peak in the optical regime, typically between the
V and B bands. They suggest a positive trend of increasing $\nu F_{\nu}$
peak flux with increasing peak frequency, in accordance with the weak
B - R vs. R correlation found above. We tested this hypothesis further
by fitting all optical SEDs with a simple parabolic shape to determine
the peak frequency, $\nu_{\rm peak}$, and the peak flux, $\nu F_{\nu}^{\rm pk}$.
The best fit values for our entire data set are plotted in Fig. \ref{nupeak}.

The $\nu F_{\nu}$ peak flux and peak frequencies are clearly correlated,
with Pearson's $r = 0.78$, with a probability for non-correlation of
$P (>r) \approx 10^{-15}$.
The best linear regression fit to the logarithms
of the $\nu F_{\nu}^{\rm pk}$ and $\nu_{\rm pk}$ values yields a power-law 
correlation $\nu F_{\nu}^{\rm pk} \propto \nu_{\rm pk}^k$ with $k = 2.05 
\pm 0.17$. A possible interpretation of this synchrotron peak shift will be
discussed in \S \ref{discussion}. 

A visual inspection of Figure \ref{nupeak} seems to suggest a steepening
of the peak flux vs. peak frequency dependence towards high peak frequencies
(and peak fluxes). However, when restricting the regression to high 
frequencies (e.g., $\nu_{\rm pk} \gtrsim 6.2 \times 10^{14}$~Hz), the
correlation between peak flux and peak frequency vanishes, therefore not
allowing for a quantification of a possible change of the correlation
slope towards high frequencies.

\begin{figure}
\plotone{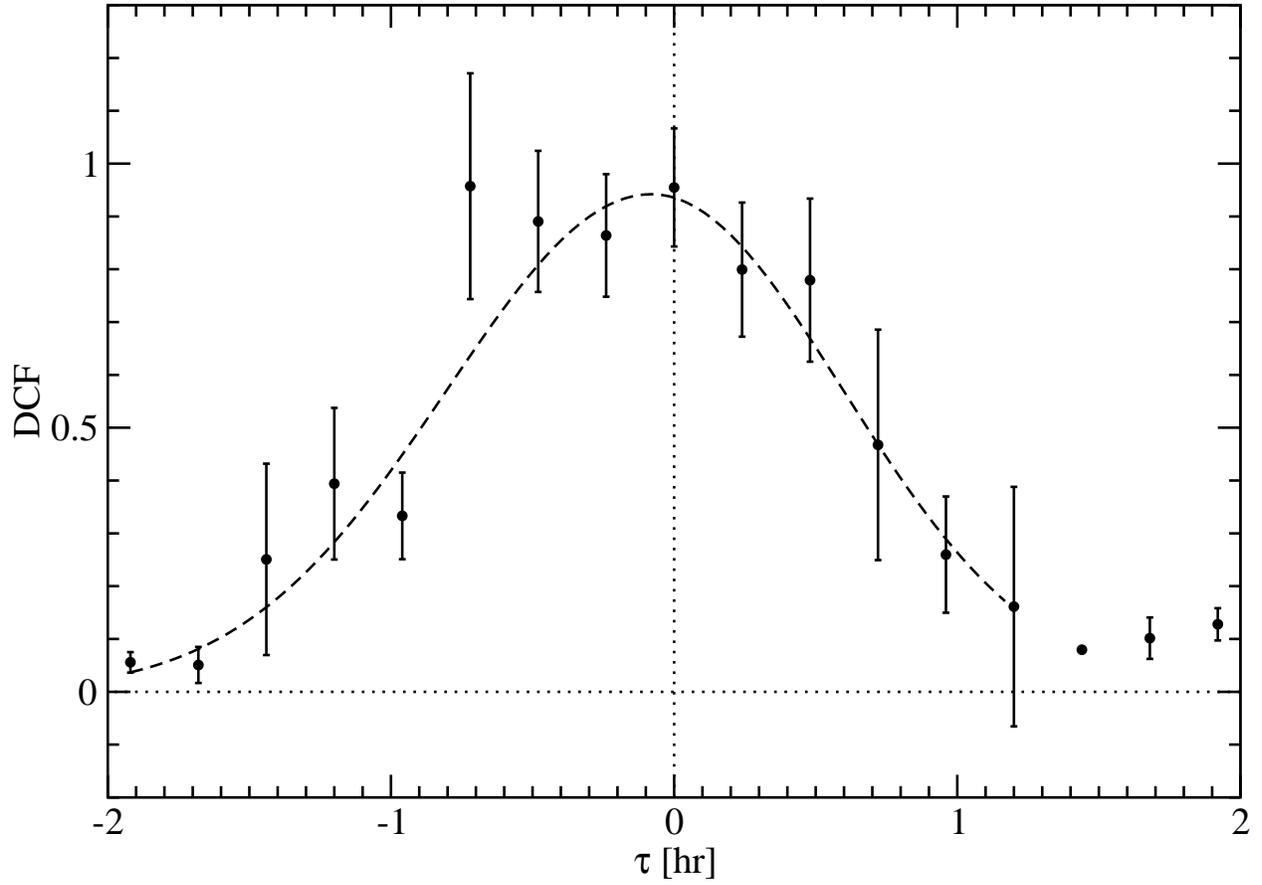}
\caption{Discrete Correlation Function between the B and R bands. A
positive $\tau$ would indicate a hard lag. The DCF has been fitted
with an asymmetric Gaussian. The best-fit peak delay is $\tau_{\rm pk}
= (-4.8 \pm 3.2)$~min.}
\label{DCFBR}
\end{figure}

\section{\label{crosscorrelations}Cross-correlation analysis}

A visual inspection of the light curves in Fig. \ref{lightcurves} and 
\ref{lc2008} suggests that the variability in all wavebands is closely 
correlated. In order to corroborate this finding, we performed a Discrete 
Correlation Function \citep[DCF,][]{ek88} analysis among all our light curves. 
Figure \ref{DCFBR} shows a typical example of the resulting DCF between
the B and R band light curves. The DCFs between the light curves of all 
bands peak at values near 1, indicating a close correlation between 
all optical bands. This is expected if the optical continuum is dominated
by nonthermal synchrotron emission of the same relativistic electron
population. Time lags between different frequency bands could potentially
serve as a diagnostic of the magnetic field strength in the emitting
region \citep[modulo the Doppler factor, see, e.g.,][]{boettcher07}. 
We therefore fitted the DCFs with an asymmetric Gaussian to determine
possible inter-band delays through the fitted peak of the DCF. However, 
any lags indicated by our DCFs are all either consistent with $0$ or 
at the $\sim 1$ -- 2~$\sigma$ level. Furthermore, our sequential
data-taking process introduces an artificial ``lag'' of up to 
$\sim 10$~min, and none of the lags found through the DCF analysis
are larger than that. Therefore, we conclude that we did not detect
any inter-band time lags.

\section{\label{discussion}Summary and Discussion}

We have presented an analysis of data from 5 years of observations
of the BL Lac object 1ES~1011+496 at the 1.3m McGraw-Hill Telescope of
the MDM Observatory. We found moderate variability on a time scale of
several days throughout most of our observations. The variability at
all (UBVRI) optical bands is well correlated with no detectable time
lags between them. The snap-shot SEDs during our observations showed
a synchrotron peak within the optical range, typically between the
V and B bands. The B - R color is 
correlated ($r = 0.57$; $P_{\rm uncorr} (>r) \approx 4 \times 
10^{-8}$)
with
the R-band magnitude, indicating a bluer-when-brighter trend. Such
a trend is observed in many BL Lac objects, where the optical emission
is strongly dominated by synchrotron emission from the jet. We note that
the opposite behaviour has been found in several quasar-type blazars,
where a slowly variable Big Blue Bump, signaling a contribution due 
to a luminous accretion disk, dilutes the continuum variability at 
the blue end of the optical spectrum \citep[e.g.,][]{raiteri08}. 

An analysis of the location of the synchrotron peak within 
the optical regime reveals a peak shift characterized by
$\nu F_{\nu}^{\rm pk} \propto \nu_{\rm pk}^k$ with $k = 2.05 \pm
0.17$, consistent with a $\nu_{\rm pk}^2$ scaling.

There is a range of possible causes for the optical (and multi-wavelength)
variability of blazar emission. These include changes in the 
Doppler factor (e.g., caused by a bending jet), injection of a 
new relativistic particle population into the jet 
(plausibly caused by a shock), a changing acceleration
efficiency (changing characteristic Lorentz factors of the radiating
electrons and possibly a change of the spectral index of the non-thermal
electron distribution), and/or a change of the magnetic field. In a
realistic scenario, several of these effects might be at work at the
same time to produce the observed blazar variability. However, one can
make simple predictions concerning the shift (in frequency and $\nu F_{\nu}$
peak flux) of the synchrotron peak for at least three cases: A changing
Doppler factor, a changing magnetic field, and a change of the
characteristic (peak) electron Lorentz factor (leaving all other
parameters of the emission region unchanged). 

The peak frequency of the synchrotron spectrum is related to a peak
in the electron spectrum at a characteristic Lorentz factor $\gamma_p$,
the magnetic field $B$ and the Doppler factor $D = \left( \Gamma \,
[1 - \beta_{\Gamma} \cos\theta_{\rm obs} ] \right)^{-1}$, where $\Gamma 
= (1 - \beta_{\Gamma}^2)^{-1/2}$ is the bulk Lorentz factor of the 
emission region and $\theta_{\rm obs}$ is the angle between the line 
of sight towards Earth and the direction of motion (the jet axis), 
through

\begin{equation}
\nu_{\rm pk} \propto \gamma_p^2 \, B \, D
\label{nupk}
\end{equation}
The $\nu F_{\nu}$ flux at the synchrotron peak is related to those
quantities through

\begin{equation}
\nu F_{\nu}^{\rm pk} \propto \gamma_p^2 \, B^2 \, D^4
\label{nuFnupk}
\end{equation}
From equations \ref{nupk} and \ref{nuFnupk}, we see that if the
variability is dominated by a changing Doppler factor, one would
expect a synchrotron peak shift as $\nu F_{\nu}^{\rm pk} \propto
\nu_{\rm pk}^4$. A change solely in the characteristic electron
Lorentz factor, $\gamma_p$, would result in a synchrotron peak
shift as $\nu F_{\nu}^{\rm pk} \propto \nu_{\rm pk}$, while a
change in only the magnetic field yields the behaviour $\nu 
F_{\nu}^{\rm pk} \propto \nu_{\rm pk}^2$.

Therefore, we conclude that the synchrotron peak shift found in
our data set is consistent with the variability being dominated
by a changing magnetic field. However, as pointed out above, we
need to caution that such a change in the magnetic field might
realistically also impact the shape of the electron distribution,
primarily through a changing synchrotron cooling time scale.
Clearly, more sophisticated analyses of this synchrotron peak
shift are needed, but are beyond the scope of this paper. 

We note that the shape of the optical SEDs found in all of our
observations contradicts the optical-UV spectrum observed by 
Swift/UVOT on 2008 May 2 and 8, which indicates a rising slope
throughout the optical regime \citep{abdo10}. However, most of our 
observations were taken during moderately faint states of the source, 
while the Swift/UVOT spectrum corresponds to a bright state, similar 
to the major optical flare that triggered the MAGIC detection 
in 2007. Given the trend of the synchrotron peak shift which 
we found in our data, it is conceivable that the Swift/UVOT 
observations correspond to an extreme case of a high synchrotron 
peak frequency, in accord with a very high optical flux.

\acknowledgments
This work was supported by NASA through Chandra Guest Observer Program 
award GO8-9100X, XMM-Newton Guest Observer Program awards NNX08AD67G
and NNX09AV45G, and Fermi Guest Investigator Program award NNX09AT82G.

\end{document}